\documentstyle[12pt,fleqn]{article}
 \catcode`\@=11

\begin{document}

\title{ Non-radial strong curvature naked singularities in
five dimensional perfect fluid self-similar space-time }

\author{Sanjay B.~Sarwe\\
 Department of Mathematics S. F. S. College,\
 \\ Seminary Hill, Nagpur-440 006, India
 \\R.V.~Saraykar\
 \\Department of Mathematics, Nagpur University Campus,\
 \\ Nagpur-440 033, India
 \\}
\date{}
\maketitle

\begin{abstract}
We study five dimensional(5D) spherically symmetric self-similar
perfect fluid space-time with adiabatic equation of state,
considering all the families of future directed non-spacelike
geodesics. The space-time admits globally strong curvature naked
singularities in the sense of Tipler
 and thus violates the cosmic censorship conjecture provided a
 certain algebraic equation has real positive roots. We further show that
it is the weak energy condition (WEC) that is necessary for
visibility of singularities for a finite period of time and for
singularities to be gravitationally strong. We, also, match the
solution to 5D Schwarzschild solution using the junction
conditions.

\end{abstract}

keywords: gravitational collapse, causal structure, naked
singularity, strength, junction condition.

\section{Introduction}

The singularity theorems of Hawking and Ellis \cite{he} have
established the fact that the gravitational collapse of
sufficiently massive star under fairly general conditions will
result in a singularity. However, these theorems do not indicate
the nature of singularity. The cosmic censorship conjecture (CCC)
articulated by Penrose \cite{pe}, in its strong version forbids
the visibility of singularity to anyone even if one is
infinitesimally close to it, whereas in the weak form, it
essentially states that gravitational collapse from a regular
initial data never creates the space-time singularity visible to
distant observers.

The development of superstring and other field theories has led to
proliferation of several articles on higher-dimensional (HD)
space-time from the viewpoint of both cosmology \cite{rs1} and
gravitational collapse \cite{hd}. Therefore, an important question
that arises: will the examples of naked singularities in
four-dimensional (4D) spherical gravitational collapse goes over
to HD space-time or not? If yes, then a related question is
whether the dimensionality of space-time has any effect on the
formation and nature of the singularity.

In recent development, it has been shown that, in null radiation
collapse \cite{gd} and in dust collapse \cite{gb}, the increase in
dimension leads to monotonic shrinkage of naked singularity
window, i.e., it favours occurrence of black holes.

              In a recent work \cite{gss}, we have
studied gravitational collapse in adiabatic self-similar fluid in
5D space-time. It turns out that the WEC is a necessary and
sufficient condition for a singularity to be globally naked and
gravitationally strong. Ghosh and Deshkar has extended these
studies to n dimensional space-time by considering null geodesics
only {\cite{sgd}}. Our purpose now is to further examine the
structure and strength of a naked singularity in the 5D perfect
fluid collapse by considering all the families of future directed
non-spacelike geodesics.

                  In Sec. 2, we determine the field equations
of 5D self-similar spherically symmetric space-time. These field
equations are used to study the occurrence of globally naked
strong curvature singularities for all the families of future
directed non-spacelike geodesics in Sec. 3. To represent a stellar
solution, the junction conditions are also discussed in Sec. 4.
The fluid solution is matched to 5D Schwarzschild solution in
retarded Eddington-Finkelstein coordinates. Conclusions are given
in Sec. 5.

\section{Self-similar perfect fluid in five dimensional space-time}
5D self-similar spherically symmetric space-time in comoving
coordinates is given by
\begin{equation}
   ds^2 = - e^{2 \nu}dt^2 +  e^{2 \psi}dr^2 + r^2 S^2 d \Omega^2.
\label{eq:me}
\end{equation}
Self similarity implies that all variables of physical interest
may be expressed in terms of the self-similarity parameter $X=t/r$.
Therefore, $\nu$, $\psi$ and $S$ are functions of $X=t/r$ only and
\begin{equation}
   d\Omega^2=d\theta^2+ \sin^2 \theta \; d \phi^2+ \sin^2 \theta
\; \sin^2 \phi \; d\vartheta^2 \label{eq:ns }
\end{equation}
is the metric on an $3$-sphere.
The pressure and energy density can be put in the form:
\begin{equation}
   P = \frac{p(X)}{8 \pi r^2}, \hspace{.5in} \rho = \frac{\eta(X)}{8
\pi r^2}. \label{edp}
\end{equation}
The stress-energy tensor for a
perfect fluid is
\begin{equation}
  T^{ab} = (\rho+P) \textbf{u}^{a} \textbf{u}^{b}+ P g^{ab} \label{eq:emt}
\end{equation}
where $\textbf{u}_a = \delta_a^t$ is the $5$-dimensional velocity.
 The self-similarity in general relativity generalizes the classical
notion of similarity and implies the existence of constants of
motion along $dX=0$, which in turn allows the reduction of the
Einstein field equations to a set of ordinary differential
equations \cite{op}:
\begin{eqnarray}
   G^t_t =- \frac{1}{S^2} + \frac{e^{-2 \psi}}{S} \Big[ X^2
\ddot{S} - X^2 \dot{S} \dot{\psi}+X S \dot{\psi} \ +\frac{(S-X
\dot{S})^2}{S} \Big]- \frac{e^{-2 \nu}}{S} \left(
\dot{S}\dot{\psi}+\frac{\dot{S}^2}{S} \right) = \frac{-\eta}{3}
\label{eq:gtt}
\end{eqnarray}
\begin{eqnarray}
   G^r_r =  - \frac{1}{S^2}+ \frac{e^{-2 \psi}}{S} \Big[- S X
\dot{\nu} + X^2 \dot{\nu} \dot{S} \
+ \frac{(S-X \dot{S})^2}{S}\Big]
 - \frac{e^{-2 \nu}}{S} \left( \ddot{S}+ \frac{\dot{S}^2}{S} -
\dot{\nu}\dot{S} \right)\ = \frac{p}{3}\label{eq:grr}
\end{eqnarray}

\begin{equation}
   G^t_r = \ddot{S}-\dot{S}\dot{\nu}- \dot{S}\dot{\psi}+\frac{S
\dot{\psi}}{X}=0
 \label{eq:gtr}
\end{equation}
where an overdot denotes the derivative with respect to $X$. We
assume that the collapsing fluid obeys an adiabatic equation of
state
\begin{equation}
   p(X) = \lambda \eta(X) \label{eos}
\end{equation}
where $0 \leq \lambda \leq 1$ is a constant. The conservation of
energy momentum tensor
\begin{equation}
   T^{ab}_{;b}=0 \label{bi}
\end{equation}
implies that
\begin{equation}
   \dot{p} + \frac{2p}{X} = - ( \eta+p)\dot{\nu}   \label{ce1}
\end{equation}
\begin{equation}
   \dot{\eta}  = - ( \eta+p)\left[\dot{\psi}+\frac{3\dot{S}}{S}
\right] \label{ce2}
\end{equation}
On integrating Eqs. (\ref{ce1}) and (\ref{ce2}), respectively, we
get
\begin{equation}
   e^{2\nu} = \gamma (\eta X^2)^{-2 \lambda/(1+\lambda)} \label{en}
\end{equation}
\begin{equation}
   e^{2\psi} =\alpha (\eta )^{-2 /(1+\lambda)}S^{-6} \label{ep}
\end{equation}
where $\alpha$ and $\gamma$ are integration constants. Eliminating
$ \ddot{S}$ from Eqs. (\ref{eq:gtt}) and (\ref{eq:grr}), we obtain
\begin{eqnarray}
   \left( \frac{\dot{S}}{S}\right)^2 V +
\frac{\dot{S}}{S}\left(\frac{\dot{V}}{2} + 3 X e^{2\nu} \right)\
+\ e^{2 \psi + 2\nu} \left(-\frac{\eta}{3}- e^{-2 \psi}+
\frac{1}{S^2} \right) =0 \label{sd}
\end{eqnarray}
and
\begin{equation}
   \dot{V} = 2 X e^{2\nu} \left[\frac{1}{3} (\eta+p) e^{2 \psi} -1
\right] = \frac{2}{3}X e^{2\nu} (H-3) \label{vd}
\end{equation}
where the quantities $V$ and $H$ are defined as follows
\begin{equation}
   V(X) = e^{2 \psi} - X^2 e^{2\nu}, \hspace{.3in} H =
(\eta+p)e^{2\psi}.
\end{equation}
We can also put $H$ as
\begin{equation}
   H = 8 \pi r^2 e^{2 \psi} \left(T^1_1 - T^0_0 \right) \label{wec}
\end{equation}
For all nonspacelike vector $V^a$, the matter satisfy weak energy
condition \cite{he} if and only if
\begin{equation}
   T_{ab} V^a V^b \geq 0 . \label{wec2}
\end{equation}
Hence, for matter satisfying weak energy condition $H(X) \geq 0$
for all $X$.

It is understood that a curvature singularity forms at the center
of the cloud ($t = 0, r=0$), where the  physical quantities like
matter density and pressure diverge. This leads to the divergence
of curvature scalars there. A singularity (a naked singularity or
a black hole) can be categorized by the existence of non-spacelike
geodesics emanating from the singularity at ($t = 0,\; r=0$). If
such geodesics exist then singularity is at least locally naked,
otherwise it is a black hole. Further, if the singularity is
naked, then there exists a real and positive value of $X_{0}$ as a
solution to the algebraic equation \cite{r1}. To analyze this, we
further study self-similar field equations.

The constants of integrations $\alpha$ and $\gamma$ can be set
equal to unity by a suitable scale transformation. Now,we define
two new functions $y= X^{\beta}$ and $U^2= {e^{2 \psi - 2
\nu}}/{X^2} = y^{-3}\eta^{-3 \beta}S^{-6}$ where $0\leq \lambda
\leq1$, $\delta=1+\lambda$ and $\beta =
{2(1-\lambda)}/{3(1+\lambda)}$. using these transformations into
Eqs.(\ref{sd}) and (\ref{vd}), we obtain
\begin{eqnarray}
   \beta y \frac{\eta'}{\eta} = \frac{1}{U^2-\lambda} \Big[
2\lambda - 3\delta \beta y U^2 \frac{S'}{S} -\frac{1}{3} \delta^2
y^3 \eta^{3 \beta/2} U^2 \Big] \label{sd1}
\end{eqnarray}
and
\begin{eqnarray}
   \left(\frac{S'}{S} \right)^2 \beta^2 y^2 (U^2-1)+
\left(\frac{S'}{S} \right) \beta y \Big[2+\frac{1}{3}\delta y^3
\eta^{3 \beta/2} U^2 \Big] -\Big[1 + \left(1-\frac{3}{\eta
S^2}\right)\frac{1}{3} y^3 \eta^{3 \beta/2} U^2 \Big]=0
\label{sd2}
\end{eqnarray}
where the prime denotes differentiation with respect to $y$. The
scale invariant quantity U represents velocity of the fluid
relative to the hypersurface $X=$ const. The case $U=1$ is related
to occurrence of naked singularity in the space-time. Therefore,
we are interested in the values of different parameters for the
solution, which take into account the case $U=1$ for some $X > 0$.
For this, we write
\begin{equation}
   \eta(X) = \eta_0 + \eta_0 \sum_{k=1}^{\infty} \eta_{k}
(y-y_0)^k \label{es}
\end{equation}
and
\begin{equation}
   S(X) = S_0 + S_0 \sum_{k=1}^{\infty} S_{k} (y-y_0)^k . \label{ss}
\end{equation}
We analyze now the solutions of above differential equations
 near the point $y= y_0= y(X_0)$ with the condition that
$U(y_0)= (\eta_0 X_0)^{-3 \beta} S_0^{-6} = 1$. On using
(\ref{es}) and (\ref{ss}), Eqs. (\ref{sd1}) and (\ref{sd2})
 take, respectively, the forms
\begin{equation}
   \eta_{1} = \frac{1}{\beta y_0 (1-\lambda)} \left(2\lambda - 3
\beta \delta y_0 S_1 - \frac{1}{3} \delta^2 y_0^3 \eta_0^{3
\beta/2} \right) \label{et1}
\end{equation}
and
\begin{equation}
   \beta y_0 S_1 = \frac{3}{6 + \delta y_0^3 \eta_0^{3
\beta/2}}\left[ 1 + \frac{1}{3} \left(1 - \frac{3}{\eta_0
S_0^2}\right) y_0^3 \eta_0^{3 \beta/2} \right] \label{ss1}
\end{equation}
Here $\eta'(y_0) = \eta_0 \eta_1$ and $S'(y_0) = S_0 S_1$.
Eliminating $S_1$ and $S_0$ from the above equations, we obtain:

\begin{eqnarray}
   Y^6 + (mn-9n)Y^4 + (6\delta - 6\lambda +9) Y^3+ 6 \delta m n Y
+27 \delta^2 - 36\lambda \delta = 0 \label{ae}
\end{eqnarray}
where
\[Y = \delta^{2/3} \eta_0^{\beta/2} y_0,\ m =\frac{\beta(1-\lambda)\eta_1}{\eta_0^{\beta-1}},
  \\n= \frac{3 \eta_0^{(\beta-2)/2}}{\delta^{2/3}} \]\
  This sixth degree algebraic equation decides
the end state of the collapse. The existence of real positive
roots of this equation will put a limitation on the physical
parameters $\eta_0$ and $\eta_1$. It is easy to verify that the
above equation can admit at the most four real positive roots. In
similar situation in four dimensions, one gets a quartic equation.

\section{ Causal Structure near the Naked Singularities}
In this section, we employ the above solution for an investigation
of the formation of a black hole or a naked singularity in a
collapsing self-similar adiabatic perfect fluid in a 5D
space-time.  A self-similar space-time is characterized by the
existence of a homothetic Killing vector:
\begin{equation}
   \xi^a = r \frac{\partial}{\partial r}+t \frac{\partial}{\partial
t} \label {eq:kl1}
\end{equation}
which is given by the Lie derivative \ $\mathcal{L}$$_{\xi}g_{ab}
=\xi_{a;b}+\xi_{b;a} = 2 g_{ab}$ where $\mathcal{L}$ is a notation
for a Lie derivative. Let $K^{i}=dx^i/dk$
 be the tangent vector to the geodesics, where $k$ is
an affine parameter. For the self similar metric  (\ref{eq:me})
the Lagrange equations immediately give
\begin{equation}
   K^{\theta}=\frac{\cos\phi\hspace{.03in} l\tan\omega}{r^2 S^2},\
K^{\phi}=\frac{\cos\vartheta\hspace{.03in} l\sin\omega}{r^2 S^2
\sin^2\theta}\ and \ K^{\vartheta}= \frac{l\hspace{.03in}
\cos\omega}{r^2 S^2 \sin^2\theta \sin^2\phi}\nonumber\
\end{equation}
 where $\omega$ and $l$ are the isotropy and impact parameters respectively.
Also,we have\
\begin{equation}
   g_{ab}K^a K^b=\mathcal{B}\label{nsc}
\end{equation}

where  $\mathcal{B}$$=0$, $\mathcal{B}$$<0$, $\mathcal{B}$$>0$
correspond to different classes of geodesics, namely, null,
timelike and spacelike, respectively.

It follows that, along non-spacelike geodesics, we have
\begin{equation}
   \xi^a K_{a} = C + \mathcal{B}\hspace{.05in}$k$ \label{eq:kl2}
\end{equation}
where $C$ is a constant. From the above algebraic equation and the
 condition (\ref{nsc}), we get
\begin{equation}
   r e^{2 \psi} K^r - t e^{2 \nu}K^t = C + \mathcal{B}\hspace{.05in}$k$ \label{kt}
\end{equation}
\begin{equation}
   - e^{2 \nu}(K^t)^2 + e^{2 \psi} (K^r)^2 +\frac{l^2}{r^2 S^2 \cos^2\omega}=\mathcal{B} \label{kr}
\end{equation}
Solving the above equations yield the following exact expressions
for $K^t$ and $K^r$:
\begin{equation}
   K^t = \frac{C \left[1+B C k\right]\left[X \pm e^{2 \psi} Q\right]}{r V(X)}\label{kt1}
\end{equation}
\begin{equation}
   K^r = \frac{C  \left[1+B C k\right]\left[1 \pm X e^{2 \nu} Q\right]}
{r V(X)} \label{kr1}
\end{equation}
where
\begin{equation}
   Q =\sqrt{e^{- 2 \psi -2 \nu}\left[1 + \frac{\left[L^2-Br^2 S^2 \cos^2\omega\right]V(X)}
{S^2 \cos^2\omega\left[1+B C k\right]^2}\right]}\nonumber
 \end{equation}
 The function $Q$ is chosen positive throughout
 and we have put $B=\frac{\mathcal{B}}{C^2}$ and $ L= \frac{l^2}{C^2}$.
  To study the nature of the singularity, we employ the technique
developed by Dwivedi and Joshi \cite{dj} by making necessary
changes for 5D case. The non-spacelike geodesics, by virtue of
Eqs. (\ref{kt1}) and (\ref{kr1}), satisfy
\begin{equation}
  \frac{dt}{dr} = \frac{X \pm e^{2 \psi} Q}{1 \pm X e^{2 \nu} Q}
\label{eq:de1}
\end{equation}
The point $t = 0, r = 0$ is a singular point of the  above
differential equation. The limiting value of $X$ reveals the exact
nature of the singularity through the analysis of non-spacelike
geodesics that terminate at the singularity:

\begin{equation}
   X_{0} = \lim_{t\rightarrow 0 \; r\rightarrow 0} X =
\lim_{t\rightarrow 0 \; r\rightarrow 0} \frac{t}{r}=
\lim_{t\rightarrow 0 \; r\rightarrow 0} \frac{dt}{dr}.
\label{eq:lm1}
\end{equation}
Using (\ref{eq:de1}) and L'H\^{o}pital's rule we get
\begin{equation}
   V(X_0)Q(X_0) = 0 \label{eq:pe}
\end{equation}
and this in turn gives
\begin{equation}
   V(X_0) = 0 \label{eq:pe3}
\end{equation}or
\begin{equation}
    Q(X_0) = 0 \ \Rightarrow L^2 V(X_0) = - S^2(X_0)\cos^2\omega\label{eq:pe7}.
\end{equation}

If Eq. (\ref{eq:pe3}) or Eq. (\ref{eq:pe7}) have any real positive
roots, then geodesics clearly terminate at the singularity with a
definite tangent, so that the central shell-focusing singularity
is at least locally naked. The smallest value of $X_0$, say
$X_0^s$, corresponds to the earliest ray escaping from the
singularity which, is called the Cauchy horizon of the space-time,
and there is no solution in the region $X < X_0^s$ \cite{r1}.
Hence, in the absence of a positive root to Eq. (\ref{eq:pe}), the
central singularity is not naked because there is no outgoing
future directed non-spacelike geodesics emanating from the
singularity. Thus, existence of the real positive roots of $
V(X_0)Q(X_0) = 0$ is a necessary and sufficient condition for the
singularity to be naked, and at least one single null geodesics in
the $(t, r)$ plane would escape from the singularity.

\subsection*{$\bullet$ Global visibility}
A naked singularity can be considered to be physically significant
singularity if a family of geodesics escape from the singularity
to far-away observers for a finite period of time. We have worked
out a detail analysis of visibility of naked singularity in
\cite{gss} and analyzed whether the singularities are globally
naked and an infinity of curves would emanate from singularity to
reach a distant observer.

\subsection*{$\bullet$ Strength}
We determine the curvature strength of a naked singularity, which
is an important aspect of a singularity \cite{ft1}.
 It is widely believed that a space-time does not admit an extension
through a singularity if it is a strong curvature singularity in
the sense of Tipler \cite{ft}. A necessary and sufficient
condition (criterion) for a singularity to be strong has  been
given by
 Clarke and Kr\'{o}lak \cite{ck}: that
for at least one non-spacelike geodesic with affine parameter $k$,
in the limiting approach to the singularity, we must have
\begin{equation}
   \lim_{k\rightarrow 0}k^2 \psi = \lim_{k\rightarrow 0}k^2 R_{ab}
K^{a}K^{b} > 0 \label{eq:sc}
\end{equation}
where $R_{ab}$ is the Ricci tensor.  We investigate the above
 condition along future-directed non-spacelike
geodesics that emanate from the naked singularity. Eq.
(\ref{eq:sc}) can be expressed as
\begin{eqnarray}
   \lim_{k\rightarrow 0}k^2 \psi = \lim_{k\rightarrow 0}\Big[ k^2
\frac{(\eta+p)C^2 e^{2 \nu}[X + e^{2 \psi} Q]^2 }{r^4 [e^{2 \psi}
- X^2 e^{2 \nu}]^2}+ \frac{\eta-2p}{2r^2}B\Big]  \label{eq:sc1}
\end{eqnarray}
Using Eqs. (\ref{kt1}), (\ref{kr1}), and L'Hospitals rule, Eq.
(\ref{eq:sc1}) for $V(X_0)=0$ turns out to be
\begin{equation}
   \lim_{k\rightarrow 0}k^2 \psi = \frac{9 H_0}{(H_0+3)^2} > 0\label{sc2}
\end{equation}
 while $Q(X_0)=0$ yields \
\begin{equation}
   \lim_{k\rightarrow 0}k^2 \psi =\frac{ H_0 U_0^{-2}}{4} > 0 .\nonumber
\end{equation}
The strong curvature condition along geodesics is satisfied if
$H_0 > 0$, which is also a necessary condition for the energy
condition. Thus, it follows that naked singularities are
gravitationally strong if the WEC is satisfied.
\section{Matching with the 5D Schwarzschild Solution}
We consider a spherical surface whose motion is described by a
timelike four-space $\Sigma$, which divides space-times into
interior and exterior manifolds.  We shall first cut the
space-time along timelike hypersurface, and then join the internal
part with the 5D Schwarzschild  solution in retarded
Eddington-Finkelstein coordinates. The metric on the whole
space-time can be written in the form
\begin{equation}
   ds^2 = \left\{ \begin{array}{ll}
        - e^{2 \nu}dt^2 +  e^{2 \psi}dr^2 + r^2 S^2 d \Omega^2, & \mbox{$r \leq r_{\Sigma}$}, \\
        - (1 -  \frac{m}{{\bf r}^2}) du^2 - 2 du d{\bf r} + {\bf r}^2 d \Omega^2, & \mbox{$r \geq  r_{\Sigma}$}.
                \end{array}
        \right.                        \label{eq:mv}
\end{equation}
The metric on the hypersurface $r=r_{\Sigma}$ is given by
\begin{equation}
   ds^2 = - d\tau^2 + {\mathcal{R}}^2(\tau) d\Omega^2. \label{bm}
\end{equation}
We suitably modify the approach of junction conditions given in
\cite{no,rw} for our 5D case. Hence, we demand
\begin{equation}
   (ds_{-}^2)_{\Sigma} = (ds_{+}^2)_{\Sigma} = (ds^2)_{\Sigma}. \label{jc1}
\end{equation}
The second junction condition is obtained by requiring the
continuity of the extrinsic curvature of $\Sigma$ across the
boundary.  This gives
\begin{equation}
   K^-_{ij} = K^+_{ij} \label{jc2}
\end{equation}
where $K^{\pm}_{ij}$ is the extrinsic curvature to $\Sigma$, given
by
\begin{equation}
   K^{\pm}_{ij} = - n_{\alpha'}^{\pm} \frac{\partial^2
x^{\alpha'}_{\pm}}{\partial \xi^i \partial \xi^j} -
n_{\alpha'}^{\pm} \Gamma^{\alpha'}_{\beta' \gamma'} \frac{\partial
x^{\beta'}_{\pm} }{\partial \xi^i} \frac{\partial x^{\gamma'}_{\pm}
}{\partial \xi^j}
 \label{ec}
\end{equation}
and where $\Gamma^{\alpha'}_{\beta' \gamma'}$ are Christoffel
symbols, $n^{\pm}_{\alpha'}$ the unit normal vectors to $ \Sigma
$, $x^{\alpha'}$ are the coordinates of the interior and exterior
space-time and $\xi^i$ are the coordinates that define $\Sigma$.
The junction condition (\ref{jc1}) yeilds
\begin{equation}
   \frac{dt}{d\tau} = \frac{1}{e^{\nu(r_{\Sigma},t)}} \label{jc3}
\end{equation}
\begin{equation}
r_{\Sigma} S(r_{\Sigma}, t) = {\bf r}(\tau) \label{jc4}
\end{equation}
\begin{equation}
   \left( \frac{du}{d\tau} \right)^{-2}_{\Sigma} = \left(1 -
\frac{m}{{\bf r}^2} + 2 \frac{d{\bf r}}{du} \right)
 \label{jc5}
\end{equation}
The non-vanishing components of intrinsic curvature $K_{ij}$ of
$\Sigma$ are determined as follows:
\begin{eqnarray}
   && K^{-}_{\tau \tau} = \left(- e^{- \psi} \nu_r \right)_{\Sigma}
\label{ecia} \\
&& K^{-}_{\theta \theta}  = \left[ e^{- \psi} r S
\left(S + r S_r \right) \right]_{\Sigma}  \label{ecib} \\
&& K^{+}_{\tau \tau} = \left[ \frac{d^2u}{d\tau^2}
\left(\frac{du}{d\tau} \right)^{-1} - \left(\frac{du}{d\tau}
\right) \frac{m}{{\bf r}^3}   \right]_{\Sigma}  \label{ecic}\\
&& K^{+}_{\theta \theta} = \left[ {\bf r} \frac{d{\bf r}}{d \tau}
 + \left(\frac{du}{d\tau} \right) \left(1 - \frac{m}{{\bf r}^2} \right) {\bf r} \right]_{\Sigma} \label{ecid}  \\
&&  K^{\pm}_{\phi \phi} = \sin^2 \theta K^{\pm}_{\theta \theta}  \label{ecie}  \\
&&  K^{\pm}_{\vartheta \vartheta} = \sin^2 \phi K^{\pm}_{\phi
\phi} \label{ecif}
\end{eqnarray}

where the subscripts $r$ and $t$ denote partial derivative with
respect to $r$ and $t$ respectively.  The unit normals to
$\Sigma$ are given by
\begin{equation}
   n_{\alpha'}^- = (0, e^{\psi(r_{\Sigma},t)}, 0, 0, 0) \label{unm}
\end{equation}
\begin{equation}
   n_{\alpha'}^+ = \left(1 - \frac{m}{{\bf r}^2} + 2  \frac{d {\bf
r}}{du} \right)^{-1/2} \left(- \frac{d {\bf r}}{du}, 1, 0, 0, 0
\right) \label{unp}
\end{equation}
From Eqs. (\ref{jc2}), (\ref{ecib}) and (\ref{ecid}) we have
\begin{equation}
   \left[ \left(\frac{du}{d\tau} \right) \left(1 - \frac{m}{{\bf
r}^2}\right) {\bf r} + {\bf r}\frac{d{\bf r}}{d\tau} \right] =
\left[ e^{- \psi} r S (S + r S_r) \right]_{\sum} \label{mf1}
\end{equation}
Using Eqs. (\ref{jc3}), (\ref{jc4}) and (\ref{jc5}), we can write
Eq. (\ref{mf1}) as
\begin{equation}
   m = r^2 S^2 \left[1 + \frac{r S_t^2}{e^{2 \nu}} - \frac{(S + r
S_r)^2}{e^{2 \psi}} \right]\label{mf}
\end{equation}
which is the total energy entrapped inside the surface $\Sigma$.
From Eqs. (\ref{ecia}) and  (\ref{ecic}), using  (\ref{jc3}), we
obtain
\begin{equation}
   \left[ \frac{d^2u}{d\tau^2} \left(\frac{du}{d\tau} \right)^{-1} -
\left(\frac{du}{d\tau} \right) \frac{m}{{\bf r}^3}
\right]_{\Sigma} = - ( e^{- \psi} \nu_r )_{\Sigma} \label{jc}
\end{equation}
Substituting Eqs. (\ref{jc3}), (\ref{jc4}) and (\ref{mf}) into
(\ref{mf1}), we get
\begin{equation}
   \left(\frac{du}{d\tau} \right)_{\Sigma}= \left[ \frac{S + r
S_r}{e^{\psi}} + \frac{r S_t}{e^{\nu}} \right]^{-1}_{\Sigma}
\label{vd1}
\end{equation}
Differentiating (\ref{vd1}) with respect to $\tau$ and using Eqs.
(\ref{mf}), we can rewrite (\ref{jc}) as
\begin{eqnarray}
    - \left(\frac{\nu_r}{e^{\psi}}  \right)_{\Sigma}= \Big[ \Big[
- \frac{r}{e^{\psi}} S_{tr} + \frac{\psi_t (S + r S_r)}{e^{\psi}}
+\frac{r \nu_t S_t}{e^{\nu}} - \frac{r S_{tt}}{e^{\nu}} - \frac{r
S_{t}^2}{e^{\nu} S} - \frac{S_t}{e^{\psi}}\nonumber\\ +
\frac{e^{\nu}}{rS} \Big(\frac{(S + r S_r)^2}{e^{2 \psi}} -1 \Big)
\Big]\times \left[ \frac{S + r S_r}{e^{\psi}} + \frac{r
S_t}{e^{\nu}} \right]^{-1} \frac{1}{e^{\nu}} \Big]_{\Sigma}.
\label{jc7}
\end{eqnarray}
Now, we translate the above equation in terms of $X = t/r$, which
gives
\begin{eqnarray}
    - \frac{1}{S^2}+ \frac{e^{-2 \psi}}{S} \left[- S X \dot{\nu} +
X^2 \dot{\nu} \dot{S}+ \frac{(S-X \dot{S})^2}{S}
\right]\
- \frac{e^{-2 \nu}}{S} \times  \left( \ddot{S}+
\frac{\dot{S}^2}{S} - \dot{\nu}\dot{S} \right)  \nonumber \\
=  \frac{e^{-\psi -\nu}(-X)}{S} \left[ \ddot{S}-\dot{S}\dot{\nu}-
\dot{S}\dot{\psi}+\frac{S \dot{\psi}}{X} \right] \label{eq:grr1}
\end{eqnarray}
Comparing (\ref{eq:grr1}) with (\ref{eq:grr}) and (\ref{eq:gtr}),
we can finally write
\begin{equation}
   (P)_{\Sigma}=0. \label{pz}
\end{equation}
Thus, the pressure vanishes at the boundary of the spherical
surface, and hence the interior 5D self-similar perfect fluid is
matched with the 5D Schwarzschild  solution.

\section{conclusions}
It has been found that the extra dimension does not affect the
occurrence of a naked singularity but rather leads to occurrence
of a strong curvature naked singularity. It is the weak energy
condition that is necessary for visibility of the singularity for
a finite period of time. The 5D spherically symmetric self-similar
space-time admits a globally strong curvature naked singularity
provided the equation $V(X)Q(X)=0$ has real positive roots, and
singularities are found to be gravitationally strong in the
Tipler's sense.

For the sake of completeness, we have matched the solution to the
5D Schwarzschild  solution in retarded Eddington-Finkelstein
coordinates, so that the resulting solution will represent the
collapse of a star.

This study generalizes the results of spherical gravitational
collapse in 4D to 5D space-time for all non-spacelike geodesics.
The formation of these naked singularities violates the cosmic
censorship conjecture. Finally, the results obtained here would
also be relevant in the context of superstring theory and for an
interpretation of how critical behaviour depends on the
dimensionality of space-time.

\end{document}